\documentclass[prl,twocolumn,superscriptaddress]{revtex4-1}

\usepackage{graphicx}  
\usepackage{bm}        
\usepackage{amssymb}   

\usepackage[usenames,dvipsnames]{color}
\usepackage[colorlinks=true, citecolor=blue, linkcolor=blue, urlcolor=blue]{hyperref}

\begin{document}
\title{Ultrafast Transient Absorption Spectroscopy of the Charge-Transfer Insulator NiO: \\
       Beyond the Dynamical Franz-Keldysh Effect}

 \author{Nicolas Tancogne-Dejean}
  \email{nicolas.tancogne-dejean@mpsd.mpg.de}
  \affiliation{Max Planck Institute for the Structure and Dynamics of Matter and Center for Free-Electron Laser Science, Luruper Chaussee 149, 22761 Hamburg, Germany}

  \author{Michael A. Sentef}
  \affiliation{Max Planck Institute for the Structure and Dynamics of Matter and Center for Free-Electron Laser Science, Luruper Chaussee 149, 22761 Hamburg, Germany}
 
 \author{Angel Rubio}
  \email{angel.rubio@mpsd.mpg.de}
\affiliation{Max Planck Institute for the Structure and Dynamics of Matter and Center for Free-Electron Laser Science, Luruper Chaussee 149, 22761 Hamburg, Germany}
\affiliation{Center for Computational Quantum Physics (CCQ), The Flatiron Institute, 162 Fifth Avenue, New York NY 10010}

\begin{abstract}
We demonstrate that a dynamical modification of the Hubbard $U$ in the model charge-transfer insulator NiO can be observed with state-of-the-art time-resolved absorption spectroscopy. Using a self-consistent time-dependent density functional theory plus $U$ computational framework, we show that the dynamical modulation of screening and Hubbard $U$ significantly changes the transient optical spectroscopy. Whereas we find the well-known dynamical Franz-Keldysh effect when the $U$ is frozen, we observe a dynamical band-gap renormalization for dynamical $U$. The renormalization of the optical gap is found to be smaller than the renormalization of $U$. This work opens up the possibility of driving a light-induced transition from a charge-transfer into a Mott insulator phase. 
\end{abstract}

\maketitle
Counterintuitive phenomena occur when a crystalline solid is exposed to time-dependent electromagnetic fields. A prominent example is the prediction by Bloch~\cite{bloch1929quantenmechanik} and Zener~\cite{jones1934h,zener1934theory} that electrons in a perfect crystals under a strong DC electric field undergoes a periodic motion instead of an uniform one, a phenomenon now referred as Bloch oscillations. A second example is the prediction of the modification of the band gap of a semiconductor under an intense electric field, resulting in a redshift of the absorption edge with the electric field, independently by Franz~\cite{franz1958einfluss} and Keldysh~\cite{keldysh1958behavior}, nowadays referred to as the Franz-Keldysh effect. Common to both phenomena is that they occur for weakly interacting electronic quasiparticles. A natural question to pose is therefore how these textbook strong-field phenomena are affected by strong electron correlations, and whether new effects emerge due to them.

Correlated quantum materials exhibit spectacular emergent phenomena already in thermal equilibrium, such as the fractional quantum hall effect \cite{tsui_two-dimensional_1982}, anomalous magnetic \cite{tokura_giant_1994} and thermal conductivities \cite{lee_anomalously_2017}, metal-insulator transitions \cite{imada_metal-insulator_1998} and high-temperature superconductivity \cite{keimer_quantum_2015}. Their out-of-equilibrium behavior, triggered by intense and ultrashort laser pulses and monitored by pump-probe spectroscopy, shows similarly diverse phenomena, such as light-induced magnetic systems\cite{kimel_nonthermal_2007, kirilyuk_ultrafast_2010, forst_driving_2011, forst_spatially_2015, mentink_ultrafast_2015, nova_effective_2017,topp2018all}, metal-to-insulator transitions\cite{rini_control_2007, caviglia_ultrafast_2012, stojchevska_ultrafast_2014}, or superconductivity\cite{fausti_light-induced_2011, kaiser_optically_2014, mitrano_possible_2016}. In laser-excited semiconductors the effects of a dynamically screened Coulomb interaction are typically taken into account self-consistently on a semi-classical level through the semiconductor Bloch equations \cite{haug_quantum_2004}. The bandgap renormalization due to the creation of charge carriers upon light excitation has been widely investigated in the past, mainly in semiconductors and insulators\cite{reynolds2000combined,nagai2004band,PhysRevB.69.205204,PhysRevLett.113.216401}. However, this simple picture of carrier-induced band-gap renormalization does not incorporate effects due to strong electronic correlations. The investigation of how strongly correlated materials react to strong laser excitations is still at a relatively early stage, and there are plenty of open questions to address.

We recently predicted that in the prototypical charge-transfer insulator NiO strong electronic correlations significantly affect the materials non-linear optical response,  highlighted by the contribution of dynamical correlation effects on the high-harmonic generation spectrum\cite{PhysRevLett.121.097402}. We found that the on-site elecron-electron interaction, parametrized by Hubbard $U$, is strongly renormalized (by about 10\%) under the presence of an intense electromagnetic field\cite{PhysRevLett.121.097402}, even when the conduction-band photodoped carriers is relatively small due to a large, factor-of-ten, mismatch between pump photon energy and optical band gap. 
Similarly model calculations revealed intriguing effects in the high-harmonic generation from a Mott insulator \cite{murakami_high-harmonic_2018} and in the band renormalization of charge-transfer insulators\cite{golez_dynamics_2018,golez_multi-band_2019}.
It is therefore legitimate to ask how the predicted renormalized on-site interaction $U(t)$ and concomitant electronic band-gap renormalization could be experimentally measured, and more importantly, if the fingerprints of a dynamical $U$ are distinguishable compared to other effects. Here, pump-probe transient absorption with sub-cycle time resolution comes to mind as a natural and unambiguous tool to probe this dynamical change in strongly correlated materials. 
Recent experimental advances have permitted to measure in GaAs and diamond the time-dependent extension of the Franz-Keldysh effect, the dynamical Franz-Keldysh effect (DFKE),  using attosecond transient absorption spectroscopy (ATAS), (see Ref.~\cite{lucchini2016attosecond} and references therein). In the DFKE, the below bandgap absorption can be viewed as a photon-assisted tunneling of the valence electrons, and as a consequence this effect follows the profile of the laser field and vanishes at the end of the laser pulse.
Using a similar procedure, the band-gap renormalization in transition metal dichalcogenides was recently measured \cite{chernikov_population_2015, pogna2016photo, meckbach_giant_2018}, demonstrating the importance of excitonic effects in the transient absorption of transition metal dichalcogenides on the observed band-gap renormalization.  
 
\begin{figure}[t]
  \begin{center}
    \includegraphics[width=\columnwidth]{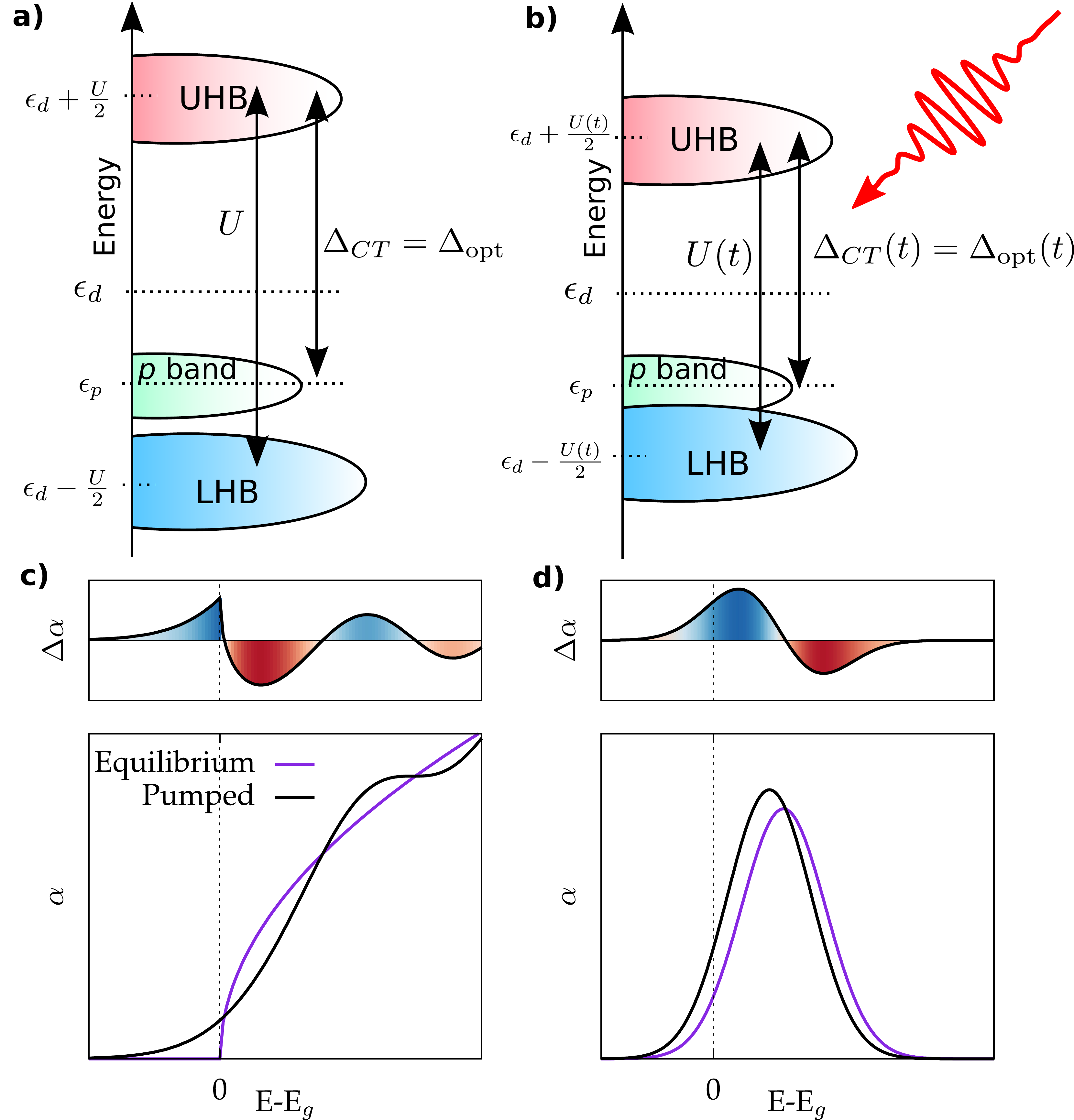}
    \caption{\label{fig:drawing} Sketch of the relative position of the transition metal $3d$ bands and the oxygen $2p$ bands for a charge-transfer insulator. Due to the on-site interaction, the $3d$ bands splits into a lower Hubbard bands (LHB) and uppper Hubbard bands (UHB). a) represents the equilibrium situation, whereas b) indicate a light-induced change of the effective electronic parameters $U$ and $\Delta_{CT}$. Bottom panel of c) shows the typical absorption $\alpha$ of a three-dimensional semiconductor (violet curve) and the absorption modified due to the Franz-Keldysh effect (black curve) \cite{Wegener_2005}. The differential absorption profile, shown in the top panel, display an increase below the bandgap, then a decrease and oscillations at higher energies. d) change in absorption and the resulting differential profile for a shift in the peak position. 
    }
  \end{center}
\end{figure}

We now turn our attention to a specific class of strongly correlated materials which is the transition metal oxides.
They are traditionally sorted as charge-transfer insulator or Mott insulator, according to a classification scheme due to Zaanen, Sawatzky, and Allen\cite{PhysRevLett.55.418}. Two effective electronic parameters are used to classify these correlated oxides. One is the Hubbard $U$ of the transition metal ions, and the other is the charge-transfer energy $\Delta_{CT}$, which represents the energy cost for promoting an electron from the $2p$ band of oxygen to the unoccupied upper Hubbard band. In the case of charge-transfer insulators with $\Delta_{CT} < U$, such as NiO, one expects the optical gap ($\Delta_{\mathrm{opt}}$) to equals $\Delta_{CT}$. This is sketched on Fig.~\ref{fig:drawing}. This raises directly the question of the dynamics of the charge-transfer gap compared to the dynamics of $U$ in a driven charge-transfer insulator. Indeed, if by using light one could make $U$ smaller than $\Delta_{CT}$, one might be able to trigger a light-induced transition from a charge-transfer insulator to a Mott insulator. Assessing the transient changes of $U$ and $\Delta_{CT}$ could in principle be resolved via time-resolved and angle-resolved photoemission spectroscopy \cite{bovensiepen_elementary_2012}. Instead, here we propose instead to use ATAS as a probe of light-induced dynamics of the optical gap, which allows us to access the dynamics of the charge-transfer gap on the attosecond timescale. 

In this Letter, we perform simulations of the transient absorption in NiO under the presence of a strong driving field. 
We drive NiO using a below band-gap excitation, which should lead to the DFKE as the strong laser will induce a polarization in the electronic system. At the same time we get a renormalized $U$. We are therefore investing here the relative importance of both effects in the transient absorption of NiO. For this, we compute its transient absorption by performing a series of time propagations. In order to get closer to the experiment, we extend our previous formalism by coupling the generalized Kohn-Sham equations to Maxwell equations, allowing to capturing the effect of the macroscopic induced transverse electric field. For each time delay, we propagate the generalized time-dependent Kohn-Sham equations of a time-dependent density functional theory plus Hubbard $U$ (TDDFT+U) framework~\cite{tancogne-dejean_self-consistent_2017} coupled to the macroscopic Maxwell equation and we later ``kick'' the system at the given time-delay to obtain the optical spectra (see Ref.~\cite{andrade_real-space_2015,castro_octopus:_2006}). The induced vector potential $\mathbf{A}_{\mathrm{ind}}(t)$ is computed from the total electronic current $\mathbf{j}(t)$ (atomic units are used throughout this Letter)
\begin{equation}
 \frac{\partial^2}{\partial t^2}\mathbf{A}_{\mathrm{ind}}(t) = \frac{4\pi c}{V}\mathbf{j}(t),
 \label{eq:maxwell}
\end{equation}
where $V$ is the volume of the cell. In order to obtain the vector potential induced by the kick, we subtract from the $\mathbf{A}_{\mathrm{ind}}(t)$ a reference calculation obtained in presence of the driving laser field but no kick~\cite{de2013simulating,walkenhorst2016tailored,de2018pump,sato2018ab}. This gives, for each time delay, the induced vector potential, starting from the out-of-equilibrium states, which is formally equivalent to performing time-dependent perturbation theory starting from the time-evolved states.
From the external electric field and the total electric field we obtain the dielectric function of light-driven NiO. 
\footnote{We found that the inclusion of the induced field results in a mild modulation of $U$ after the end of the laser pulse, but neither to a further sizable reduction of $U$ nor a change in its dynamics.}
Note that this approach is an alternative to the one used in Refs.~\onlinecite{lucchini2016attosecond,sato2018ab}.

All the calculations presented here were performed for bulk NiO, which is a type-II antiferromagnetic material below its N\'eel temperature ($T_N=523$K\cite{cracknell_space_1969})\footnote{We neglected the small rhombohedral distortions and considered NiO in its cubic rock-salt structure, which does not affect the result of calculated optical spectra. Calculations were performed using norm-conserving pseudo-potentials, a lattice parameter of 4.1704~\AA\, a real-space spacing of $\Delta r=0.293$ Bohr, and a $16\times16\times8$ $\mathbf{k}$-point grid to sample the Brillouin zone.}.
The driving field is taken along the [001] crystallographic direction in all the calculations. We consider a laser pulse of 7.5 fs duration (FWHM), with a sin-square envelope for the vector potential. The carrier wavelength $\lambda$ is 800\,nm, corresponding to a carrier photon energy of 1.55\,eV, which is much smaller than the calculated bandgap of 4.14\,eV. We checked that similar results are obtained for below bandgap excitation with a much longer wavelength of 2100\,nm~\cite{SI}.
The time-dependent wavefunctions, electric current, and $U_{\mathrm{eff}}$ are computed by propagating generalized Kohn-Sham equations within real-time TDDFT+U, as provided by the Octopus package~\cite{andrade_real-space_2015,castro_octopus:_2006}.
We employed the PBE functional~\cite{perdew_generalized_1996} for describing the semilocal DFT part, and we computed the effective $U_{\mathrm{eff}}=U-J$ for the O $2p$ ($U^{2p}_{\mathrm{eff}}$) and Ni $3d$ orbitals ($U^{3d}_{\mathrm{eff}}$), using localized atomic orbitals from the corresponding pseudopotentials~\cite{tancogne-dejean_self-consistent_2017}.
All calculations are propagated for 15 fs after the kick, to avoid spurious numerical effects. A Gaussian broadening of $\eta=0.3$ eV is added to mimic the experimental broadening. Here we considered a moderate intensity of $I_{\mathrm{ext}}= 5.0\,$TW.cm$^{-2}$ for the external field in matter, corresponding to an intensity in matter of $I_{\mathrm{mat}}\sim 0.4\,$TW.cm$^{-2}$.\footnote{This value corresponds to the intensity of the external field, defined without the refractive index. The total field acting on electrons is reduced (due to the local-field effect, i.e., macroscopic induced field) by a factor of $\epsilon_1^2$, where $\epsilon_1 \sim5.4$ is the dielectric constant at 800nm of bulk NiO at equilibrium. The intensity in matter is thus estimated to be of $I_{\mathrm{mat}}= 0.4\,$TW.cm$^{-2}$, this time taking into account the refractive index in the definition of the intensity.}\\
%
The time-dependent generalized Kohn-Sham equation within the adiabatic approximation reads \footnote{The nonlocal part of the pseudopotential is omitted for conciseness}
\begin{eqnarray}
 i\frac{\partial}{\partial t}|\psi^\sigma_{n,\mathbf{k}}(t)\rangle = \Big[\frac{(\hat{\mathbf{p}}-\mathbf{A}_{\mathrm{tot}}(t)/c)}{2} + \hat{v}_{\mathrm{ext}} + \hat{v}_{\mathrm{H}}[n(\mathbf{r},t)]\nonumber\\
 + \hat{v}_{\mathrm{xc}}[n(\mathbf{r},t)] + \hat{V}_{U}[n(\mathbf{r},t),\{n^{\sigma}_{mm'}\}]\Big]|\psi_{n,\mathbf{k}}^\sigma(t)\rangle,
\end{eqnarray}
where  $|\psi_{n,\mathbf{k}}^{\sigma} \rangle$ is a Bloch state with a band index $n$, at the point $\mathbf{k}$ in the Brillouin zone, and with the spin index $\sigma$, $\hat{v}_{\mathrm{ext}}$ is the ionic potential, $\mathbf{A}_{\mathrm{tot}}(t)$ is the total vector potential containing the induced one of Eq.~\ref{eq:maxwell},  $\hat{v}_{\mathrm{H}}$ is the Hartree potential, $\hat{v}_{\mathrm{xc}}$ is the exchange-correlation potential, and $\hat{V}^\sigma_{U}$ is the (non-local) operator,
\begin{equation}
\hat{V}^\sigma_{U}[n,\{n^{\sigma}_{mm'}\}] = U_{\mathrm{eff}}\sum_{m,m'}( \frac{1}{2}\delta_{mm'} - n^{\sigma}_{mm'} ) \hat{P}_{m,m'}^\sigma\,.
\label{eq:pot_V_U}
\end{equation}
Here $\hat{P}_{mm'}= |\phi_{m}^\sigma\rangle\langle\phi_{m'}^\sigma|$ is the projector onto the localized subspace defined by the localized orbitals $\{\phi_{m}^\sigma\}$, and $n^{\sigma}$ is the density matrix of the localized subspace. 
The expressions of $U$ and $J$ 
can be found for instance in Ref.~\cite{agapito_reformulation_2015}.

\begin{figure}[t]
  \begin{center}
    \includegraphics[width=\columnwidth]{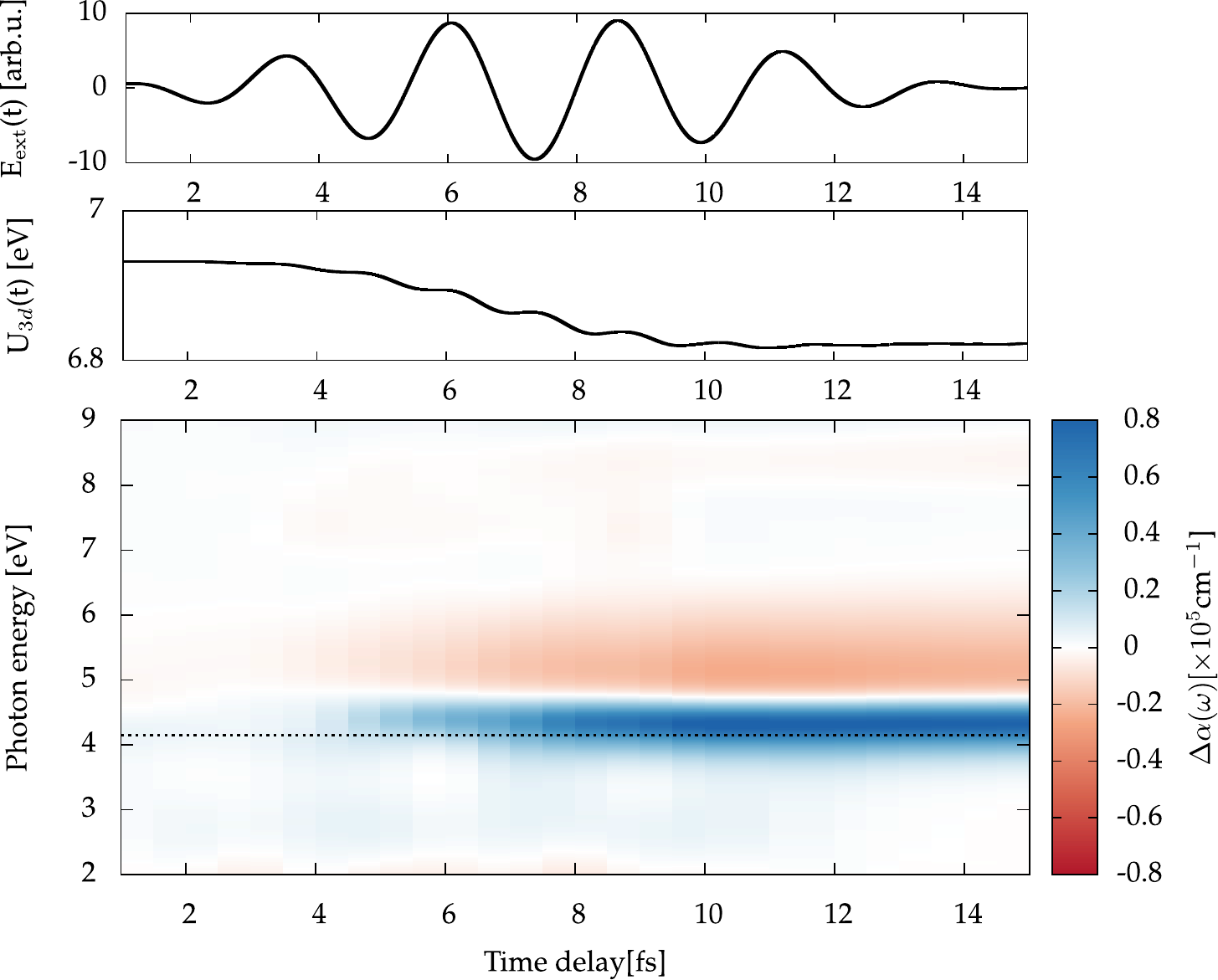}
    \caption{\label{fig:TransAbs} Calculated transient absorption of NiO for a 800\,nm driving field. The top panel shows the time profile of the external driving electric field $\mathbf{E}_{\mathrm{ext}}(t)$. The middle panel shows the corresponding time-evolution of the on-site $U$ for the $3d$ orbitals of Ni. A similar variation is obtained for the $2p$ orbitals of oxygen atoms. The bottom panel shows the calculated transient absorption of NiO. The dashed line indicates the calculated ground-state gap of NiO.}
  \end{center}
\end{figure}

Then, we compute the ATAS of bulk NiO under a strong driving field, focusing on the visible range. The driving vector potential is shown in the top panel of Fig.~\ref{fig:TransAbs}, and the corresponding dynamics of $U$ is shown in the middle panel  of Fig.~\ref{fig:TransAbs}. The calculated transient absorption of NiO is shown in the bottom panel. Our results show a main feature around the band gap of the NiO in its groundstate, indicated by the horizontal dashed line.
The increase of the spectral weight at lower energy, together with the decrease of spectral weight at higher energy, form a differential profile, characteristic of a shift in energy, here towards lower energy (as expected from a shift in the main absorption peak position, from Fig.~\ref{fig:drawing}d). Clearly, our simulations predict band-gap renormalization, in agreement with the reduction of the Hubbard $U$ due to a change in the electronic screening\cite{PhysRevLett.121.097402}. It is now interesting to come back to the DFKE effect. Indeed, in the picture of the DFKE, the band gap of the material also changes due to the presence of an external driving field.
However, in the DFKE, the change in the gap follows the profile of the driving field and disappears at the end of the pulse. This is in clear contrast with the present results, in which the band-gap renormalization, similar to the light-induced change of $U$, remains intact at the end of the pulse. 

\begin{figure}[t]
  \begin{center}
    \includegraphics[width=\columnwidth]{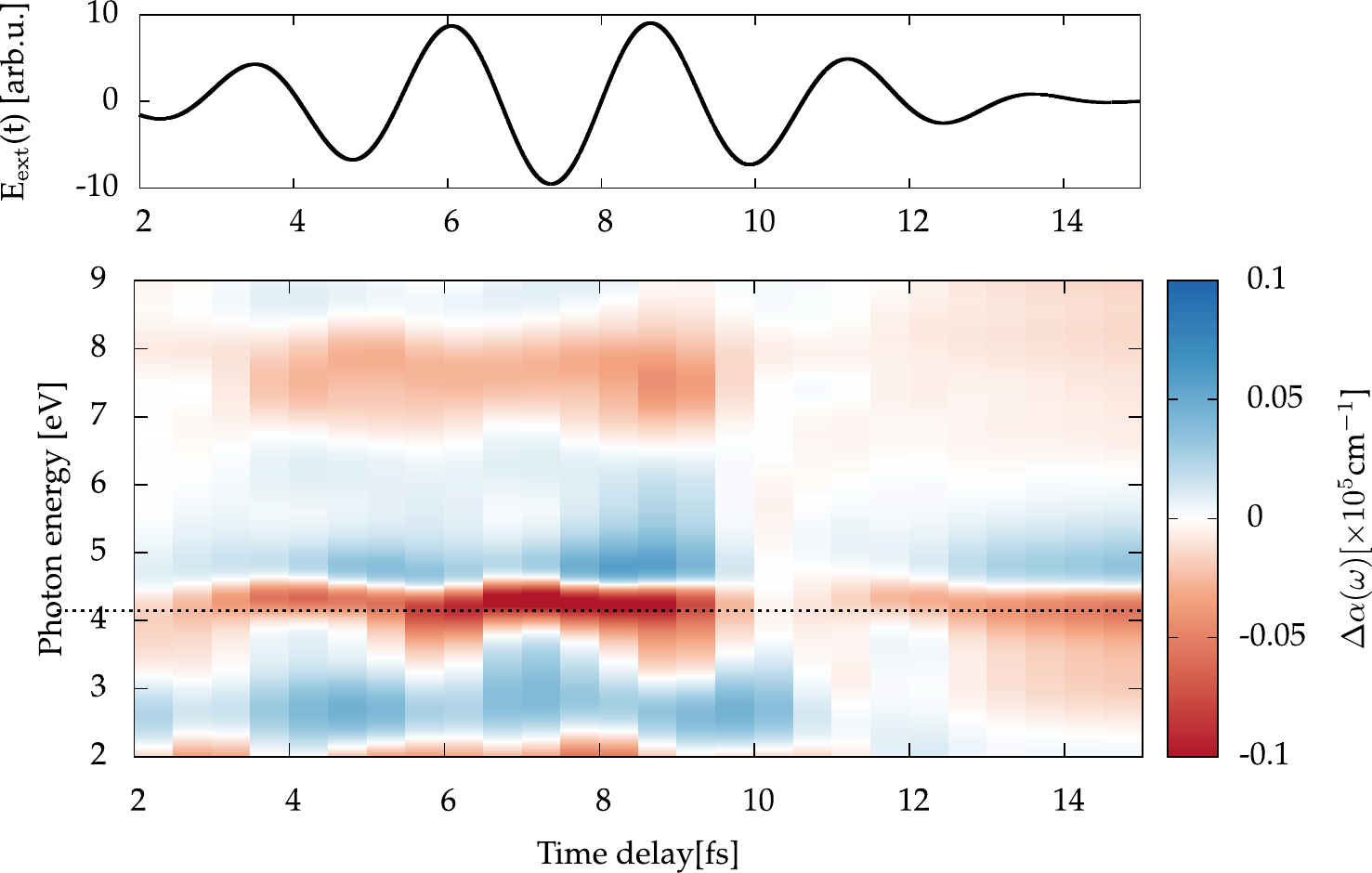}
    \caption{\label{fig:TransAbs_Freeze} The same as Fig.~\ref{fig:TransAbs}, but keeping $U$ fixed to its ground-state value of 6.93\,eV during the time evolution. }
  \end{center}
\end{figure}

To understand the difference between our results and the DFKE picture, we perform same calculations, but keeping the $U$ frozen during the whole time evolutions. It is important to note that the changes in electronic populations are fully taken into account, as well as local-field effects and all the local inhomogeneities that would be accounted for in a usual TDDFT calculation based on a (semi)local functional. However, effects due to the change in  $U$, and hence the changes in the electronic screening, are not accounted for. Put differently, we suppress effects due to dynamical correlations. Our results for fixed $U$ are shown in Fig.~\ref{fig:TransAbs_Freeze}. 
In clear contrast with Fig.~\ref{fig:TransAbs}, we do not observe a unique differential profile over the entire range of time delays, but rather three different features: i) an increase of the absorption below the bandgap, ii) a decrease of the absorption starting at the position of the bandgap, iii) an increase of absorption at higher energy. These features can be easily understood in terms of the DFKE, as presented in Fig.~\ref{fig:drawing}c. Moreover, these changes vanish at the end of the laser pulse. The extracted dynamics of the optical gap is shown in Supplemental Information~\cite{SI}.
Comparing the scales of the change in the absorption, one observes that the change in absorption for a fixed $U$ is smaller than for a dynamical $U$ by a factor of 8. 

This gives us some clear insight on how dynamical correlation effects modify the DFKE. Indeed, in the case of a dynamical $U$, the dominating feature is the band-gap renormalization, which could either originate from the dynamical $U$ or from the usual dynamical screening from light-induced charge carriers. As we show below, we found that the band-gap renormalization originate here mostly from the dynamical $U$.
Importantly, we find the transient change in the absorption induces by the exponential tail acquired by the wavefunctions in the solid under a strong field, which is the DFKE, appears here to be almost negligible, as no clear below-bandgap absorption is observed.
This result shows that the band-gap renormalization related to the dynamical $U$ has a very strong effect on the transient absorption of NiO, revealing the possibility to measure it experimentally.

It is important to note that the modification of the absorption spectrum at the end of the laser pulse is of different nature in the case of dynamical or frozen $U$. In the case of frozen $U$, the change in the absorption spectrum comes from the carrier-induced broadening and Pauli-blocking effects. However, as we are considering here a non-resonant, below-bandgap excitation of moderate intensity, these effects, which are directly related the excited carrier density, are not so important here. In the case of the dynamical $U$, the main effect comes from the band-gap renormalization, as $U$ gets dynamically renormalized, which is a consequence of the change in dielectric screening.  Due to the reduction of the gap, the excitation of carriers is also increased in the case of light-reduced $U$, confirmed by a reduction of the Ni atom magnetization is observed~\cite{PhysRevLett.121.097402}, and the charge dynamics is also changed.

We now come back to the classification of transition-metal oxides by Zaanen, Sawatzky, and Allen. In their classification, Mott insulators and charge-transfer insulators differ by the fact that in the former case, one has $U<\Delta_{CT}$ with charge-transfer energy $\Delta_{CT}$, whereas in the latter case, the charge transfer energy $\Delta_{CT}$ is smaller than $U$. 
The possibility of engineering the effective electronic parameters on the time scale of the optical cycle could lead to fundamentally very exciting applications, such as light-driven Mott insulators. However, this requires not only $U$ to reduce, but also $\Delta_{CT}$ to become larger than the on-site interaction. From the result of Fig.~\ref{fig:TransAbs}, it is clear that the charge-transfer gap $\Delta_{CT}$ also gets reduced by the light pulse.

\begin{figure}[t]
  \begin{center}
    \includegraphics[width=0.95\columnwidth]{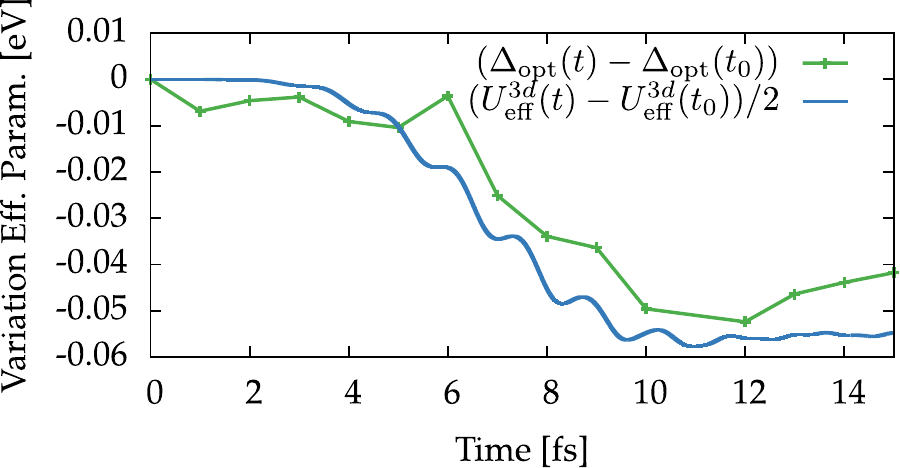}
    \caption{\label{fig:Diagram} Calculated variation of the optical gap, extracted from the calculated transient absorption, that is compared to the variation of $U_{\mathrm{eff}}$. }
  \end{center}
\end{figure}

It is therefore interesting to ask the following question: How does the light pulse steer NiO in the Zaanen-Sawatzky-Allen phase diagram? 
To answer this question, we first extract from our simulations the value of the charge-transfer gap. As this is not a well-defined observable, we use instead the optical gap, extracted directly from the transient absorption spectra (details in the Supplemental Material\cite{SI}).
The extracted optical gap is shown in the top panel of Fig.~\ref{fig:Diagram}, and compared with the time profile of Hubbard U.

Clearly, $U$ reduces faster than the optical gap $\Delta_{\mathrm{opt}}$ (and therefore than $\Delta_{CT}$). This result implies that these two quantities are not renormalized equally, a condition necessary to drive a charge-transfer insulator toward the Mott phase. Our calculations indicate that the optical gap closely follow the dynamics of $U^{3d}_{\mathrm{eff}}(t)/2$, which is consistent with the usual picture of charge-transfer insulation (as sketched in Fig.~\ref{fig:drawing}), in which the optical gap is formed by O $2p$ bands and transition metal $3d$ bands, with only the latter affected by the change in $U$. Deviations originate from carrier-induced effects such as Pauli-blocking and carrier-induced broadening.

In summary, we investigated the ATAS in NiO using TDDFT+U simulations coupled with macroscopic Maxwell equations. We found that for a prototypical strongly-correlated material, NiO, the picture of the dynamical Franz-Keldysh effect has to be reconsidered due to correlation effects. Indeed the dominant feature in the transient absorption trace becomes the band-gap renormalization, whereas when we freeze the $U$, we recover the transient absorption structures characteristic of the dynamical Franz-Keldysh effect.
From our results, we extracted the time profile of the optical gap, showing that while it also gets renormalized by the light, it decreases only half as fast compared to $U$, namely like $U_{\mathrm{eff}}(t)/2$, as expected from the point of view of a charge-transfer insulator. 
Our results are general, applicable to other kind of correlated insulators, and open the door to driving a charge-transfer insulator to the Mott-Hubbard phase using intense femtosecond laser fields.

\begin{acknowledgments}
This work was supported by the European Research Council (ERC-2015-AdG694097), and the Flatiron Institute, which is a division of the Simons Foundation. M.A.S. acknowledges financial support by the DFG through the Emmy Noether programme (SE 2558/2-1). 
We would like to thank O. D. M\"ucke for fruitful discussions.
\end{acknowledgments}

%

\end{document}